**Correcting a SHAPE-directed RNA structure by a mutate-map-rescue approach**

Siqi Tian,[1] Pablo Cordero,[2] Wipapat Kladwang,[3] and Rhiju Das[1, 2, 3,] *

[1]Department of Biochemistry, [2]Biomedical Informatics Program, and [3]Department of Physics, Stanford University, Stanford, California 94305, United States

*Corresponding Author

**ABSTRACT**

The three-dimensional conformations of non-coding RNAs underpin their biochemical functions but have largely eluded experimental characterization. Here, we report that integrating a classic mutation/rescue strategy with high-throughput chemical mapping enables rapid RNA structure inference with unusually strong validation. We revisit a paradigmatic 16S rRNA domain for which SHAPE (selective 2´-hydroxyl acylation with primer extension) suggested a conformational change between apo- and holo-ribosome conformations. Computational support estimates, data from alternative chemical probes, and mutate-and-map ($M^2$) experiments expose limitations of prior methodology and instead give a near-crystallographic secondary structure. Systematic interrogation of single base pairs via a high-throughput mutation/rescue approach then permits incisive validation and refinement of the $M^2$-based secondary structure and further uncovers the functional conformation as an excited state (25±5% population) accessible via a single-nucleotide register shift. These results correct an erroneous SHAPE inference of a ribosomal conformational change and suggest a general mutate-map-rescue approach for dissecting RNA dynamic structure landscapes.

**INTRODUCTION**

RNA plays critical roles in diverse cellular and viral processes ranging from information transfer to metabolite sensing to translation (Nudler and Mironov 2004; Amaral et al. 2008; Zhang et al. 2009; Zhang et al. 2010; Breaker 2012). For the most complex of these processes, RNAs must adopt and interconvert between specific three-dimensional structures (Zhang et al. 2010), but for the vast majority of systems, these conformations remain experimentally uncharacterized. In particular, current prediction methods do not yet offer clear metrics of statistical confidence, probe the possibility of multiple conformations (including weakly populated 'excited' states), or provide routes to cross-validation through independent experiments. The situation is particularly problematic since RNAs can form multiple alternative secondary structures whose helices are mutually exclusive (Nudler and Mironov 2004; Henkin 2008; Haller et al. 2011).

One approach for probing large numbers of RNA molecules involves chemically modifying RNA and reading out these events via electrophoresis or deep sequencing (Mitra et al. 2008; Lucks et al. 2011; Pang et al. 2011; Yoon et al. 2011). Numerous reagents, including protein nucleases (Walczak et al. 1996; Grover et al. 2011; Siegfried et al. 2011), alkylating chemicals such as dimethyl sulfate (DMS) (Wells et al. 2000; Tijerina et al. 2007; Cordero et al. 2012a), and hydroxyl radicals (Adilakshmi et al. 2006; Das et al. 2008; Ding et al. 2012), have been leveraged to modify or cleave RNA in a structure-dependent manner. Protection of nucleotides from modification, typically signaling the formation of base-pairs, can guide manual or automatic secondary structure inference (Mathews et al. 2004; Mitra et al. 2008; Vasa et al. 2008). Strong cases have been made for using reagents that covalently modify 2´-hydroxyls followed by read out via primer extension (SHAPE) (Merino et al. 2005; Mortimer and Weeks 2007; Deigan et al. 2009; Watts et al. 2009; McGinnis et al. 2012) and then applying these data as pseudo-energy bonuses in the free-energy minimization algorithms, such as RNAstructure (Mathews et al. 2004; Reuter and Mathews 2010; Hajdin et al. 2013).

Assessing the accuracy of structure mapping methods is becoming a major issue as these approaches are being applied to large RNA systems – such as entire cellular transcriptomes – for which crystallographic, spectroscopic, or phylogenetic methods cannot be brought to bear (Kertesz et al. 2010; Underwood et al. 2010; Kladwang et al. 2011a). In some cases, the resulting models have disagreed with accepted structures (Quarrier et al. 2010; Kladwang et al. 2011c; Hajdin et al. 2013; Sükösd et al. 2013), motivating efforts to estimate uncertainty (Kladwang et al. 2011c; Ramachandran et al. 2013), to incorporate alternative mapping strategies (Cordero et al. 2012a; Ding et al. 2013; Kwok et al. 2013) and to integrate mapping with systematic mutagenesis (mutate-and-map, $M^2$) (Kladwang and Das 2010; Kladwang et al. 2011a; Kladwang et al. 2011b; Cordero et al. 2013). Even these improvements do not provide routes to validation through independent experiments, and structure inferences remain under question (Wenger et al. 2011).

A powerful technique for validating RNA structures is the rescue of disruptive perturbations from single mutations through compensatory mutations that restore Watson-Crick pairs. Mutation/rescue has typically been read out through catalytic reactions such as self-cleavage (Wu and Huang 1992; Macnaughton et al. 1993). For RNAs without well-established functional assays, chemical mapping offers a potentially general readout for mutation/rescue, but has not been previously explored. Here, we demonstrate the applicability and power of such high-throughput mutation/rescue for the 126-235 RNA, a 16S ribosomal domain and S20-protein-binding site (Brodersen et al. 2002) for which a prior SHAPE study (Figure 1) proposed a large-scale conformational change (Deigan et al. 2009). Our new analyses and data instead gave a model that was consistent with the RNA's accepted secondary structure up to a single-nucleotide register shift. In-depth analysis of high-throughput mutation/rescue experiments further revealed that the crystallographic conformation was present as an 'excited state' at 25% of the population. The resolution and in-depth validation of these results on the 126-235 RNA suggest that this study's mutate-map-rescue approach could be a generally powerful pipeline for interrogating structures and excited states of non-coding RNA domains.

**RESULTS**

*Reproducibility and robustness of SHAPE modeling*

Before applying newer approaches, we carried out standard SHAPE chemical mapping on the 126-235 RNA read out by capillary electrophoresis (CE). Our SHAPE data with the acylating reagent 1-methyl-7-nitroisatoic anhydride (1M7) (Mortimer and Weeks 2007), averaged across four replicates, agreed with previously published data (Deigan et al. 2009) for the 126-235 RNA in the context of the full ribosome and in different flanking sequences (Figure 1D). In particular, sequence assignments agreed with prior work, and were additionally verified by co-loading chemical mapping samples with dideoxy sequencing ladders and measurements using the Illumina-based MAP-seq (Multiplexed Accessibility Probing read out through sequencing) protocol (Seetin et al. 2013) (SI Figure S1A-B). In ribosome crystals and phylogenetic analyses, the 126-235 RNA domain forms a junction of four helices (Zhang et al. 2009) (Figure 1B); while the prior SHAPE-directed model (referred to hereafter as the '1D-data-guided model') proposed two different helices (Figure 1A). The SHAPE data were consistent with both the ribosome crystallographic secondary structure as well as with the prior 1D-data-guided secondary structure. In both models, the nucleotides observed to be SHAPE-reactive either do not form Watson-Crick base pairs or are next to such nucleotides.

We then carried out secondary structure predictions guided by these SHAPE data, using the RNAstructure modeling algorithm with default SHAPE pseudo free energy parameters, which were derived from prior 16S rRNA analysis (Deigan et al. 2009). The resulting model (Figure 2A) was identical to the 1D-data-guided model inferred in the previous study and distinct from the crystallographic secondary structure (Figure 1B). Hereafter, we refer to the helical segments in the crystallographic secondary structure as P1a-c, P2a-b, P3, and P4a-b, since they provide a finer level of description than the conventional 16S rRNA helix numbering (H122, H144, H184, H198). The 1D-data-guided model retains P1a-c, P2b, and P3, but adds an additional helix P2* and forms distinct helices that we label as alt-P1d, alt-P4, and alt-P5 (Figure 2A).

Confusion regarding SHAPE protocols and recent reports of artifacts by us and others (Wilkinson et al. 2006; Kladwang et al. 2011b; Leonard et al. 2013) motivated us to further test the robustness of our experimental SHAPE protocol and algorithm. We obtained results that were the same, within estimated experimental error, upon varying folding solution conditions (Wilkinson et al. 2006; Deigan et al. 2009; Kladwang et al. 2011b), acylating reagents (1M7 and NMIA) (Merino et al. 2005; Mortimer and Weeks 2007), reverse transcription conditions (Mills and Kramer 1979), quantitation software (Yoon et al. 2011; Karabiber et al. 2013), normalization schemes (Deigan et al. 2009), and RNAstructure modeling software versions (Mathews et al. 2004; Hajdin et al. 2013) (Figure 1D, SI Figures S1A-B, S2). Overall, these data confirmed the reproducibility of the SHAPE experimental method and modeling procedure across different conditions and by different groups.

*Non-parametric bootstrapping gives low confidence estimates*

Procedures for estimating uncertainties in SHAPE-directed modeling have not yet become widely accepted. We recently proposed that useful helix-by-helix support values might be calculated by a resampling procedure called non-parametric bootstrapping (Efron et al. 1996; Kladwang et al. 2011c). This conceptually simple procedure has found wide use in complex statistical problems such as phylogenetic inference in which parametric models for likelihood or posterior probabilities are unavailable or untrustworthy. Mock datasets are 'bootstrapped' from the experimental dataset by resampling with replacement from the collection of data-derived energy bonuses. These data sets mimic scenarios in which data at particular residues might be missing or extra data at particular residues are available (e.g., from multiple-probe methods). The data are then input into the same secondary structure prediction algorithm. The frequencies at which helices arise in these bootstrap replicates provide 'bootstrap supports'. Low support values indicate that alternative structural models exist and are nearly as consistent with the experimental data as the original model. While this procedure has been criticized as being overly conservative (Ramachandran et al. 2013), experimental benchmarks on noncoding RNAs of known structure (Kladwang and Das 2010; Kladwang et al. 2011a; Kladwang et al. 2011b) and simulation-based studies (ST, RD, unpublished results) confirm that bootstrap supports provide numerically accurate indicators of helix confidence.

For the 126-235 RNA SHAPE modeling, bootstrapping calculations gave a wide range of values for the 1D-data-guided model (Figure 2). On one hand, helices that were shared between the 1D-data-guided model and the crystallographic model gave bootstrap supports above 80% (see P1a, P1c, P2b, and P3 in Figure 2A). On the other hand, helices that were rearranged in the 1D-data-guided model gave lower bootstrap supports (43%, 78%, 38%, and 13% for alt-P1d, P2*, alt-P4, and alt-P5 respectively). In contrast to prior suggestions (Ramachandran et al. 2013), the strengths of these supports did not simply reflect helix length. For example, the short P3 (three base pairs) attained a high support of 83%, while the second-longest helix alt-P1d attained a lower support of 43%. Instead, the helices with low supports lay in regions which could form numerous alternative structures calculated to have energies nearly as low as the final 1D-data-guided model, including the crystallographic secondary structure (blue arrows, Figure 2B).

The low bootstrap supports for the rearranged helices, as well as the difficulty of discriminating between multiple alternative structures with the available data, motivated us to acquire measurements on the 126-235 RNA that might validate or falsify their existence. Independent information from alternative modifiers highlighting bases whose Watson-Crick edges are available for alkylation (A-N1 or C-N3, by DMS) (Tijerina et al. 2007) or for the carbodiimide reaction (G-N1 or U-N3, by CMCT) (Walczak et al. 1996; Cordero et al. 2012a) were obtained, and secondary structure prediction guided by these data show similarly low bootstrap values (SI Figure S1D-E). Except for non-canonical pairings in P2*, this analysis provided weak or no support to the 1D-data-guided model above and suggested that confidently disambiguating the RNA's solution structure would require methods with higher information content.

### *Two-dimensional mutate-and-map ($M^2$) resolves ambiguities in modeling*

Compared to the 'one-dimensional' (1D) structure mapping approaches above, a recently developed two-dimensional (2D) expansion of chemical mapping offered the prospect of higher confidence modeling of the 126-235 RNA. The mutate-and-map method ($M^2$) involves the parallel synthesis of separate constructs harboring single mutations at each nucleotide of the RNA (SI Table S2), followed by chemical mapping of each construct at nucleotide-resolution (Kladwang and Das 2010; Kladwang et al. 2011a; Kladwang et al. 2011b). Observation of an initially protected region that becomes reactive upon mutation of a sequence-distant region provides evidence for pairing between the two regions. Mutations that are unique in their effect and that do not release nucleotides other than their partner appear as punctate features in the $M^2$ data; such signals provide the strongest evidence for nucleotide-nucleotide interactions (Kladwang et al. 2011b).

$M^2$ electropherograms for the 126-235 RNA were acquired as in prior work (Kladwang et al. 2011b) (Figure 3A). Several features provided consistency checks in the $M^2$ analysis. Perturbations near the site of the mutations were visible as a diagonal feature from top-left to bottom-right. In addition, in several expected regions, mutations led to punctate features corresponding to 'release' of sequence-distant nucleotides. For example, G138 was exposed by C225G and no other mutation. These data, as well as nearby features (marked I in Figure 3A) supported P1c. Similarly, punctate features of increased reactivity at G168 by C153G and G184 by C193G defined hairpin P2b and P3 (marked II and III, respectively, on Figure 3A). These segments, P1c, P2b, and P3, were present in the all models above, including the prediction by RNAstructure with no experimental data (SI Figure S1C).

Additional punctate features provided discrimination between the 1D-data-guided secondary structure and other models. First, C217G released nucleotide G200 (IV in Figure 3A). The only other mutations that perturbed G200 were changes near this nucleotide in sequence and G187C, which caused a change in SHAPE profile throughout the RNA, presumably reflecting a major rearrangement of secondary structure. This feature supported pairing of C217 and G200, which occurs in P4a of the crystallographic secondary structure; it also disfavored the 1D-data-guided model, in which these nucleotides are instead partnered with different nucleotides (C217-G145 and G200-U208). Second, G146C released C176, while affecting no other region of the 126-235 RNA (V). This feature supports base-pairing of G146-C176, which occurs in crystallographic P2a but not in the 1D-data-guided model. Additional punctate features suggested interactions between G220 and A143 (VI) and A174 and A199 (VII). These features do not connect nucleotides that are Watson/Crick paired but may form non-canonical interactions.

To fully integrate the $M^2$ data into a structural model, we carried out automated secondary structure prediction with RNAstructure (Mathews et al. 2004). This analysis takes into account the single-nucleotide-resolution features as above but also leverages additional, less punctate features that correspond to, e.g., disruption of multiple base pairs upon mutations. The weights of these features are calculated as Z-scores (Figure 3B), which down-weight any regions that are highly variable across constructs (see Methods). Consistent with visual analysis above, the resulting secondary structure (referred to hereafter as the '2D-data-guided model'), recovered helices P1a-c, P2b, P3, and the non-canonical P2* with high confidence (support values of 96% or greater) (Figure 3C-D). Furthermore, this 2D-data-guided model agreed with the crystallographic secondary structure in helix P4b (support value 99%), which is entirely absent from the 1D-data-guided structure. Minor discrepancies with the crystallographic secondary structure occurred in P2a and in P4a. Some bootstrap replicates

(36%) recovered two base pairs in helix P2a, but a larger fraction of replicates (63%) returned this helix with the pairings shifted by one nucleotide, a secondary structure we called shift-P2a (compare Figure 1C and 1B). Nevertheless, the overall support totaled over P2a and shift-P2a was strong (98%). Similarly, bootstrap replicates sampled alternative registers for the P4a helix (shift-P4a and the crystallographic P4a at 85% and 13%, respectively, summing to 98%). In contrast, only 2% of bootstrap replicates gave secondary structures consistent with the alt-P4 rearrangement in the 1D-data-guided model (Figure 2B, 3D).

As an independent confirmation of the $M^2$ analysis, we repeated the $M^2$ experiments with the SHAPE modifier 1M7 (instead of NMIA) and again with the DMS modifier. Automated secondary structure prediction guided by these separate $M^2$ datasets returned models indistinguishable from the NMIA-based analysis above, with similar residual ambiguities in the registers of P2a and P4b (SI Figure S3). Overall, the $M^2$ analysis recovered the crystallographic secondary structure, up to potential single-nucleotide register shifts in helices P2a and P4b.

*Systematic falsification of 1D-data-guided model through mutation/rescue*

To more deeply interrogate and test the 126-235 RNA's secondary structure, we sought to validate or falsify base pairs in the models above through mutation/rescue experiments. A classic technique for validating RNA models is the rescue of disruptive perturbations from single mutations through compensatory mutations. Rescue of function in double mutants predicted to restore Watson-Crick pairs disrupted by separate single mutations provides strong evidence for base-pairing of those nucleotides. Mutation/rescue approaches are well developed for RNA molecules whose structure is coupled to a functional readout, such as self-cleavage (Wu and Huang 1992; Macnaughton et al. 1993). But such a general and systematic validation method has not been explored for cases without well-established functional assays. Here, chemical mapping offers a single-nucleotide-resolution readout of perturbation and restoration of structure (Kladwang et al. 2011b). The crystallographic model, 1D-data-guided and 2D-data-guided models provided specific pairing hypotheses to test. The same synthesis pipeline (Kladwang et al. 2011b) leveraged to synthesize 96 single mutants for $M^2$ measurements permitted facile synthesis of these additional RNA variants (SI Table S2). Figure 4 shows the capillary electropherograms of pairings in question as well as quantitated data.

Discriminating the secondary structure models required identification of nucleotides with different pairings across models. As an example, G201 paired with C207 within the alt-P4 helix of the 1D-data-guided model, but with C217 in the shift-P4a helix of the 2D-data-guided model. Single mutation of G201C resulted in clear changes in SHAPE data over an approximately 20-nucleotide region (180 to 198; Figure 4F). The 1D-data-guided model predicted that mutation C207G would restore this pairing. However, the double mutant G201C/C207G retained the disruptions observed in the G201C single mutant as well as additional changes observed in the C207G single mutant (Figure 4F). An analogous experiment on the same pairing but different mutations (G201U/C207A) gave similar results, with no observed rescue (Figure 4E).

In contrast to the above experiment, the 2D-data-guided model predicted that a different mutation, C217G, would restore the pairing disrupted by the G201C mutation. Such a rescue was indeed observed in G201C/C217G (Figure 4S). The effect was even more striking given that the rescuing C217G mutation produced dramatic disruptions throughout the entire RNA when implemented as a single mutant (Figure 4S). The SHAPE reactivities for the entire double mutant RNA were indistinguishable from the wild type RNA with a minor exception: weak reactivity at G204 in the wild type was suppressed in the double mutant. Together with similar observations with G201U/C217A (Figure 4R), these experiments provided strong evidence for shift-P4a predicted by the 2D-data-guided model.

In addition to the pairing options G201 above, we tested additional base-pairings by mutation/rescue studies on alt-P1d, alt-P4, and alt-P5 (Figure 4A-O). None of these experiments exhibited disruptions by single mutants that were rescued by double mutants (SI Figure S4A, S4E). Three double mutants (A199U/U209A, A143U/U219A and A143C/U219G) gave reactivity profiles similar to wild type, but the single mutations alone did not introduce significant disruptions either (Figure 4A, 4K, 4L).

For the 2D-data-guided model, mutation/rescue resolved ambiguities in the P2a/shift-P2a and P4a/shift-P4a regions. (Mutation/rescue for helix P4b was rendered difficult by its low stability; see SI Figure S4D.) For the pairings of shift-P2a, double mutants did not exhibit rescue of disruptions observed in single mutants (Figure 4W-

Z). Therefore, another set of double mutants was designed based on the register of helix P2a seen in the crystallographic model (Figure 4AG-AJ). In each of these cases, the double mutants did rescue the perturbations induced by single mutants, restoring the SHAPE reactivity seen in the wild type RNA. (In G145U/G177A, the SHAPE reactivity of G177 was suppressed, consistent with replacement of the original non-canonical G/G pair with a canonical A/U pair.) Taken together, these data strongly favored the formation of the crystallographic P2a.

In the case of shift-P4a, as with G201C/C217G above, all double mutants rescued disruptions observed in single mutants, giving strong evidence for helix shift-P4a (Figure 4P-V). Furthermore, the quantitated data showed strong agreement with each other (SI Figure S4B, S4F).

*Quantitative dissection of the P4a helix register shift*

The data above strongly favored the crystallographic secondary structure. The only exception was a subtle one: a single-nucleotide register shift in helix P4a, which was experimentally validated through mutation/rescue of the shifted pairings. We sought to further probe this register shift by carrying out mutation/rescue experiments based on crystallographic P4a but not in shift-P4a (Figure 4AA-AF). Surprisingly, the double mutants SHAPE data showed rescue of disruptions induced by single mutations, and led to SHAPE data indistinguishable from the wild type data, except for stronger reactivity at G204 and the disappearance of weak reactivity at G198 (SI Figure S4C). These data strongly supported the presence of the crystallographic helix P4a, in apparent contradiction to the results above supporting the 2D-data-guided helix shift-P4a.

The paradox of mutation/rescue data supporting both registers could be resolved if 126-235 RNA interconverts between P4a and shift-P4a registers at equilibrium. The double mutants chosen to test shift-P4a perturbed the equilibrium and favored the shift-P4a structure; and alternative stabilization occurred for the P4a double mutants. This model provided testable predictions: double mutants should converge to two slightly distinct SHAPE reactivity profiles depending on which register they preferentially stabilized. Furthermore, the wild-type SHAPE profile should be decomposable into a linear combination of the P4a-stabilized and shift-P4a-stabilized profiles.

Indeed, the double mutants observed to stabilize P4a gave distinct SHAPE reactivity data from those that stabilized shift-P4a at nucleotides G204 (increased) and G198 (decreased); see Figure 4AA-AF (blue arrows), SI Figure S4B-C, S4F-G. In accord with the model, these are the two nucleotides brought into and out of the helix, respectively, by the register shift; compare Figure 4S and 4AD. Finally, the wild-type data could be fitted to an equilibrium mixture of the shift-P4a-stabilized and P4a-stabilized profiles, as determined from the double mutants (Figure 5A). The equilibrium fractions of shift-P4a and P4a were fitted to be 75±5% and 25±5%, corresponding to a free energy difference of 0.7±0.2 kcal/mol. In the context of the full ribosome, this equilibrium may be perturbed to favor the crystallographic/phylogenetic structure by RNA tertiary contacts or S20 protein binding, providing possible checkpoints for 16S rRNA assembly that can be tested in future experiments (SI Figure S6).

**DISCUSSION**

*Towards validated models of RNA structure from chemical mapping*

As a test case for single-nucleotide-resolution chemical mapping, we have carried out an in-depth dissection of the solution structure of a model system with complex RNA structure, a 110-nucleotide segment of the 16S ribosomal RNA that was previously suggested to undergo a large-scale conformational rearrangement between its intrinsic solution structure and its state within the assembled small ribosomal subunit. Through a simple computational procedure, we first discovered that prior SHAPE-directed modeling (Deigan et al. 2009) as well as DMS and CMCT chemical mapping measurements did not carry sufficient information content to determine the 126-235 RNA's secondary structure. We then leveraged a more information-rich technology, the mutate-and-map ($M^2$) approach, to infer a more confident model, which disagreed with prior approaches and was instead identical to the structure observed in ribosome crystallography, up to small ambiguities in helix register (Zhang et al. 2009). Finally, we demonstrated that a rich array of mutation/rescue experiments could be designed, carried out, and interpreted to provide strong tests and refinement of these RNA structure models. In addition to resolving the secondary structure at nucleotide resolution, these data revealed the exact crystallographic secondary structure, which is shifted from the dominant structure by a single-nucleotide register change, to be present as an

'excited state'. The overall mutate-map-rescue ($M^2R$) pipeline offers numerous advantages over prior methodology.

*Limitations of previous mutational analysis*

Prior mutational analysis was not designed to precisely validate or falsify base pairs and instead gave data that did not discriminate between possible models. One mutant, M1, harbored two mutations U208C and U218C, and was speculated to stabilize the 1D-data-guided model by switching two U-G pairs to more stable C-G pairs. Although the overall 1D SHAPE profile of M1 is indistinguishable from wild-type (SI Figure S5A-B), this observation does not exclude alternative pairing schemes. Mutation U208C is also consistent with both the 2D-data-guided and crystallographic model since U208 is unpaired in both models. Mutation U218C can result in pairing with another G in both 1D-data-guided and 2D-data-guided models: G144 in alt-P1d and G200 in shift-P4a. In our final model (Figure 5B), forming a G200-C218 pair would lock the structure in shift-P4a and give G204 reactivity lower than the wild type RNA. This slight perturbation was indeed observed in M1 (SI Figure S5G). Thus, the data for M1 is consistent with both prior and current models.

The second mutant, M2, was designed to stabilize the crystallographic model with two mutations G145C and U219C. A perturbed SHAPE profile at G177 was taken to support a conformational change (Deigan et al. 2009). Nevertheless, the observed changes can be explained by small differences in the structures. G145C stabilizes the original non-canonical G145-G177 pair in crystallographic model into a Watson/Crick base-pair, reducing the reactivity of G177 (SI Figure S5H, S6A). U219C alters a U-G pair to C-G pair in P4a, thus stabilizing and also locking it in P4a conformation, enhancing the reactivity of G204 and A205 in M2 (blue arrow, SI Figure S5H). The mutations render the M2 sequence unable to form base-pairs seen in the 1D-data-guided model (SI Figure S5E-F). The M2 data again do not discriminate between previous and new models.

In summary, the variants tested in the prior study, M1 (U208C/U218C) and M2 (G145C/U219C), were double mutants but did not probe compensatory mutations; those results were consistent with numerous structures, including the new model here. More generally, the number of alternative models for an RNA system is vast and difficult to enumerate; to our knowledge, only compensatory rescue provides incisive mutational discrimination amongst secondary structures.

*Excited states*

Nearly all experimental RNA structural methods, ranging from solution chemical mapping to crystallography, focus on inferring the dominant RNA conformation in solution. Obtaining an experimental description of weakly populated states necessarily requires more information and has typically involved mutations to trap such states (Pan and Woodson 1998; Russell et al. 2002; Gilbert et al. 2007). The mutate-map-rescue ($M^2R$) approach herein involved a systematic panel of such mutations and provided a clear view of one such excited state. Several different sets of mutations (Figure 4P-V, 4AA-AF) in the region 199-203 converged to a chemical reactivity profile that was distinct from the wild type. The effects of single mutations and the reactivity perturbations could be explained by formation of the crystallographic P4a conformation, which is shifted in register by a single nucleotide from the dominant solution conformation shift-P4a. This register shift model was additionally confirmed by double-mutant analysis and by demonstrating the recovery of the wild type reactivity from a linear combination of measurements on mutants stabilizing the conformations. Notably, our approach is analogous to – but significantly faster than – recent mutation-coupled NMR methods for inferring and stabilizing excited states of RNA model systems (Dethoff et al. 2012).

Here, the 'excited' state of the 126-235 RNA, present at 25% population, has the register observed in crystallography of the entire 16S rRNA (Figure 5). Within the full ribosome, this functional helix register may be stabilized by tertiary interactions with the rest of the 16S rRNA or with ribosomal proteins (SI Figure S6B-C). Indeed, the 1D SHAPE profile of full-length 16S rRNA from previous study resembles P4a-stabilized mutants at G204 (Figure 1D). An A-minor interaction between G203/C214 and A465 is observed in crystallographic structure (SI Figure S6B), suggesting that this tertiary contact may stabilize the P4a conformation in full-length 16S rRNA context. The small ribosomal protein S20 can also bind the 126-235 region, and it plays a crucial role in stabilizing this region of the 5´ domain (Rydén-Aulin et al. 1993; Brodersen et al. 2002). These scenarios suggest a novel checkpoint for ribosome assembly based on locking the P4a/shift-P4a register shift, and should be resolvable through future experiments. More generally, the mutate-map-rescue pipeline holds promise for discovering and validating excited states for other RNAs, especially if the inference of a structure ensemble from the available data can be fully automated.

*A general mutate-map-rescue ($M^2R$) pipeline*

This study has delineated an expansion of conventional chemical mapping that enables systematic inference, testing, and refinement of RNA structure domains, including the correction of prior misleading inferences and possibility of inferring excited states. While unusually detailed for a chemical mapping study, the entire mutate-map-rescue pipeline described herein was carried out with the same commercially available reagents, equipment, and synthesis strategy as our standard high-throughput chemical mapping protocol (Kladwang et al. 2011b; Lucks et al. 2011). With current technologies, the presented pipeline is generally applicable to non-coding RNA domains up to 300 nucleotides in length. We therefore propose that this mutate-map-rescue approach can be generally adopted as a 'best practice' for finalizing RNA models inferred from chemical mapping.

**METHODS**

**RNA synthesis**

Double-stranded DNA templates were prepared by PCR assembly of DNA oligomers with maximum length of 60 nt ordered from IDT (Integrated DNA Technologies). DNA templates contain a 20-nt T7 RNA polymerase promoter sequence (TTCTAATACGACTCACTATA) on the 5´ end, followed by sequence of interest. One hairpin with single-stranded buffering region was added on both ends to flank the region of interest. A 20-nt Tail2 sequence (AAAGAAACAACAACAACAAC) was put on the 3´ end (SI Table S1). Primer assembly scheme for all constructs was designed by an automated MATLAB script (NA_Thermo, available at https://github.com/DasLab/NA_thermo) (Kladwang et al. 2011b).

PCR reactions, consisted of 200 pmol of terminal primers and 2 pmol of internal primers were carried out as previously described (Kladwang et al. 2011b). PCR products were purified using Ampure XP magnetic beads

(Agencourt) on a 96-well Greiner microplate format following manufacturer's instructions. DNA concentration were measured by UV absorbance on Nanodrop 1000 spectrophotometer (Thermo Scientific). DNA templates were verified by sequencing (PAN core facility, Stanford University). *In vitro* transcription reactions were described previously (Kladwang et al. 2011b), followed by same purification and quantification steps as DNA. Sequences and purities of RNA samples were confirmed by reverse transcription in presence of each ddNTP.

**Chemical modification**

One-dimensional chemical mapping, mutate-and-map ($M^2$), and mutation/rescue were carried out in 96-well format as described previously (Kladwang et al. 2011a; Kladwang et al. 2011b; Cordero et al. 2013). Prior to chemical modification, 1.2 pmol of RNA was heated up and cooled to remove secondary structure heterogeneity (90 °C for 2 min and cooled on ice for 2 min) and folded for 20 min at 37 °C in 15 µL of one of the following buffers: 10 mM $MgCl_2$, 50 mM Na-HEPES, pH 8.0 (our standard) (Kladwang et al. 2011b); 5 mM $MgCl_2$, 200 mM KOAc, 50 mM Na-HEPES, pH 8.0 (Deigan et al. 2009); 10 mM $MgCl_2$, 100 mM NaCl, 50 mM Na-HEPES, pH 8.0 (Wilkinson et al. 2006); and 50 mM Na-HEPES, pH 8.0.

RNA was modified by adding 5 µL of modification reagent [0.5% dimethyl sulfate (DMS) prepared by mixing 1 uL 10.5 M DMS into 9 µL ethanol, and then 190 µL water; 21 mg/mL 1-cyclohexyl-(2-morpholinoethyl) carbodiimide metho-p-toluene sulfonate (CMCT); 5 mg/mL 1-methyl-7-nitroisatoic anhydride (1M7); or 12 mg/mL N-methylisatoic anhydride (NMIA)] (Merino et al. 2005; Mortimer and Weeks 2007; Tijerina et al. 2007; Cordero et al. 2012a). $ddH_2O$ or anhydrous DMSO was used as background control. Modification reactions were incubated at room temperature for 20 min and then quenched appropriately (5 µl of 0.5 M Na-MES, pH 6.0 for SHAPE and CMCT or 2-mercaptoethanol for DMS). All modifiers were made fresh before use. Quenches also included 1 µL of poly(dT) magnetic beads (Ambion) and 0.065 pmols of FAM-labeled Tail2-A20 primer ($A_{20}$-GTTGTTGTTGTTGTTTCTTT) for reverse transcription. Samples were separated and purified using magnetic stands, washed with 100 µL 70% ethanol twice, and air-dried. Beads were resuspended in 2.5 µL ddH2O for reverse transcription and 2.5 µL reverse transcription mix, then incubated at 55 °C for 30 min. RNAs were degraded by adding 5 µL 0.4 M NaOH and incubating at 90 C for 3 min. Solutions were cooled down on ice then neutralized with 5 μL acid quench (1.4 M NaCl, 0.6 M HCl, and 1.3 M NaOAc). Fluorescent labeled cDNA was recovered by magnetic bead separation, rinsed twice with 40 μL 70% ethanol and air-dried. The beads were resuspended in 10 μL Hi-Di formamide (Applied Biosystems) with 0.0625 μL ROX-350 ladder (Applied Biosystems) and eluted for 20 min. Supernatants were loaded to capillary electrophoresis sequencer (ABI3100). Sequencing ladders were prepared analogously, without any chemical modification but with inclusion of each 2´-3´-dideoxy-NTP (ddNTP) equimolar to each dNTP during reverse transcription.

To verify sequence assignments in downstream analysis, an additional "co-loaded" sample composed of each sequencing ladder (5 μL) and cDNA derived from SHAPE-probed RNA (5 μL) was also measured. RNA chemical mapping using an alternative deep-sequencing readout was carried out analogously to the method above but included an additional ligation step to permit Illumina sequencing (Seetin et al. 2013).

**Data Processing and structural modeling**

The HiTRACE software package version 2.0 was used to analyze CE data (MATLAB toolbox is available at https://github.com/hitrace (Yoon et al. 2011), a web server is also available at http://hitrace.org (Kim et al. 2013)). Electrophoretic traces were aligned and baseline subtracted using linear and non-linear alignment routines as previously described (Kim et al. 2009). Sequence assignment was accomplished manually with verification from sequencing ladders and the co-loaded samples. Band intensities were obtained by fitting profiles to Gaussian peaks and integrating.

Rigorous normalization, correction for signal attenuation, and background subtraction were enabled by inclusion of referencing hairpin loop residues (GAGUA) at both 5´ and 3´ ends, 10x dilution replicates, and no-modification controls (get_reactivities in HiTRACE). Briefly, values for saturated peaks were obtained from 10x dilutions. Signal attenuation was corrected from 5´ to 3´ ends based on the relative reactivity between 5´ and 3´ referencing hairpin loop intensities. Reactivities of all the chemical profile were normalized against GAGUA. For most comparisons, the average of reactivities in each of the two GAGUA hairpins were set to two for best comparison with previously reported data (Deigan et al. 2009).

Data-driven secondary structure models were obtained using the Fold program of the RNAstructure package (Mathews et al. 2004; Reuter and Mathews 2010). For secondary structure models guided by 1D chemical mapping, pseudo-energy slope and intercept parameters of 2.6 kcal/mol and –0.8 kcal/mol (RNAstructure version 5.4) (Mathews et al. 2004) or, where stated, 1.8 kcal/mol and -0.6 kcal/mol (version 5.5) (Hajdin et al. 2013) were used. To obtain $M^2$-guided secondary structure models, Z score matrices for $M^2$ datasets were calculated as previously described (Kladwang et al. 2011b). Briefly, Z scores were calculated for each nucleotide reactivity by subtracting the average reactivity of this nucleotide across all mutants and dividing by standard deviation (output_Zscore_from_rdat in HiTRACE). $M^2$ seeks to identify release of putative base pair partners of a nucleotide upon mutation; therefore, negative Z scores and positions with high average reactivity (cutoff is 0.8) were excluded. Background subtraction and signal attenuation correction were not applied to $M^2$ data, since Z scores are independent of those steps (which would otherwise introduce noise). Z score matrices were used as base-pair-wise pseudo-energies with a slope and intercept of 1.0 kcal/mol and 0 kcal/mol (Kladwang et al. 2011b).

Helix-wise confidence values were calculated via bootstrapping as described previously: mock datasets were generated by sampling the mutants with replacement and comparing the helices of the resulting mock-data-driven models with those in the model obtained using the full data (Kladwang et al. 2011b). An independent analysis using the QuSHAPE software was performed following given instructions (Karabiber et al. 2013). In all modeling steps, full-length sequences (including flanking elements) were used for prediction.

**Structural equilibrium fitting**

Equilibrium fractions of each structure were determined by assuming that shift-P4a-stabilized mutants and P4a-stabilized mutants double mutants completely stabilize the register-shifted and crystallographic structure respectively – their reactivity profiles therefore represent the reactivity profile for each state. The reactivites of nucleotide G204 from shift-P4a-stabilized double mutants and P4a-stabilized double mutants were taken to fit the wild-type reactivity by a linear combination ratio.

**Structural visualization**

Secondary structure images were generated by VARNA (Darty et al. 2009). Atomic model of crystallographic data from PDB entry 3I1M (Zhang et al. 2009) was visualized in PyMol (The PyMOL Molecular Graphics System, Version 1.5.0.4 Schrödinger, LLC.). Non-canonical base-pairing and long-range interactions were mapped in assistance with RNA 3D Hub (Petrov et al. 2013). Secondary structure diagram in Leontis/Westhof nomenclature (Leontis and Westhof 2001) was drawn in Illustrator (Adobe).

**Data Reposition**

All chemical mapping datasets, including one-dimensional mapping, mutate-and-map, and mutation/rescue, have been deposited at the RNA Mapping Database (http://rmdb.stanford.edu) (Cordero et al. 2012b) under the following accession codes: 16S_STD_0001, 16S_NMIA_0001, 16S_1M7_0001, 16S_DMS_0001, and 16S_RSQ_0001.

**ACKNOWLEDGMENT**

We acknowledge financial support from a Stanford Graduate Fellowship (S.T.), a Conacyt fellowship (P.C.), the Burroughs Wellcome Foundation (CASI to R.D.), and NIH R01 R01GM102519. We thank members of the Das lab for comments on the manuscript.

**LEGENDS**

**Figure 1. 126-235 RNA models and M$^2$ data.** *(A-C)* Secondary structure of 1D-data-guided, 2D-data-guided, and crystallographic model. Helix color codes are shared for all figures. *(D)* Normalized SHAPE reactivity of 126-235 RNA measured herein and from prior study (Deigan et al. 2009). Standard deviations (SD) are shown, N = 7.

**Figure 2. 1D-data-guided SHAPE (1M7) results of 126-235 RNA.** *(A).* Secondary structure prediction using 1-dimensional SHAPE data. Nucleotides are colored with SHAPE reactivities. Difference from crystallographic model is drawn in yellow/gray lines. Percentage labels give bootstrap support values. *(B).* Bootstrap support matrix of *(A).* Blue arrows point at alternative conformations.

**Figure 3. 2D-data-guided SHAPE (NMIA) results of 126-235 RNA.** *(A).* Mutate-and-map (M$^2$) dataset probed by NMIA. Structural features (I-V) highlight evidence of base-pairings and interactions. *(B).* Z-score contact-map extracted from *(A). (C-D).* Secondary structure prediction and bootstrap support matrix using 2-dimensional M$^2$ data (NMIA). Percentage labels give bootstrap support values. Difference from crystallographic model is drawn in yellow/gray lines.

**Figure 4. Mutation/rescue results of 126-235 RNA.** Electropherograms of mutation/rescue to test base-pairings from 1D-data-guided (alt-P1d, alt-P4 and alt-P5) *(A-O)*, 2D-data-guided (shift-P2a and shift-P4a) *(P-Z)*, and crystallographic (P2a and P4a) *(AA-AJ)* models. Blue arrows mark G204 in P4a-stabilized mutants.

**Figure 5. Equilibrium fitting and final model.** *(A)* Linear fitting of P4a/shift-P4a equilibrium based on G204. An enlarged view of G204 reactivities is on the right. Standard deviations (SD) are shown, N = 5. *(B)* Model of 126-235 RNA secondary structure in solution, including P4a /shift-P4a dynamics.

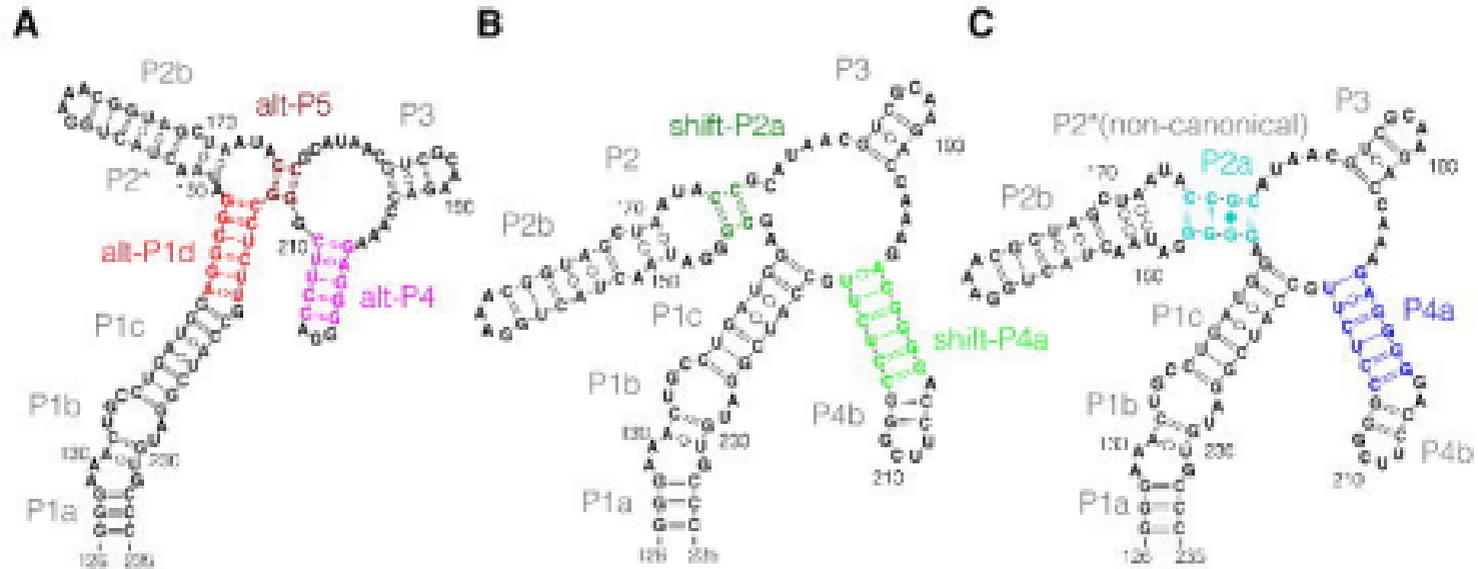

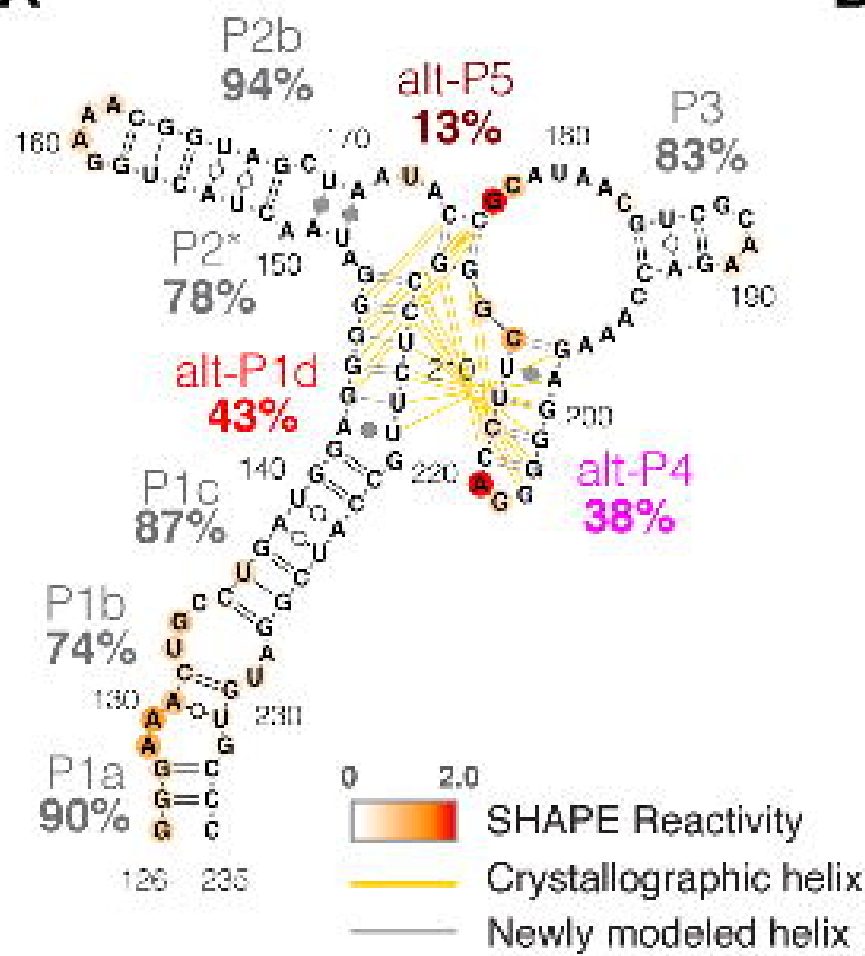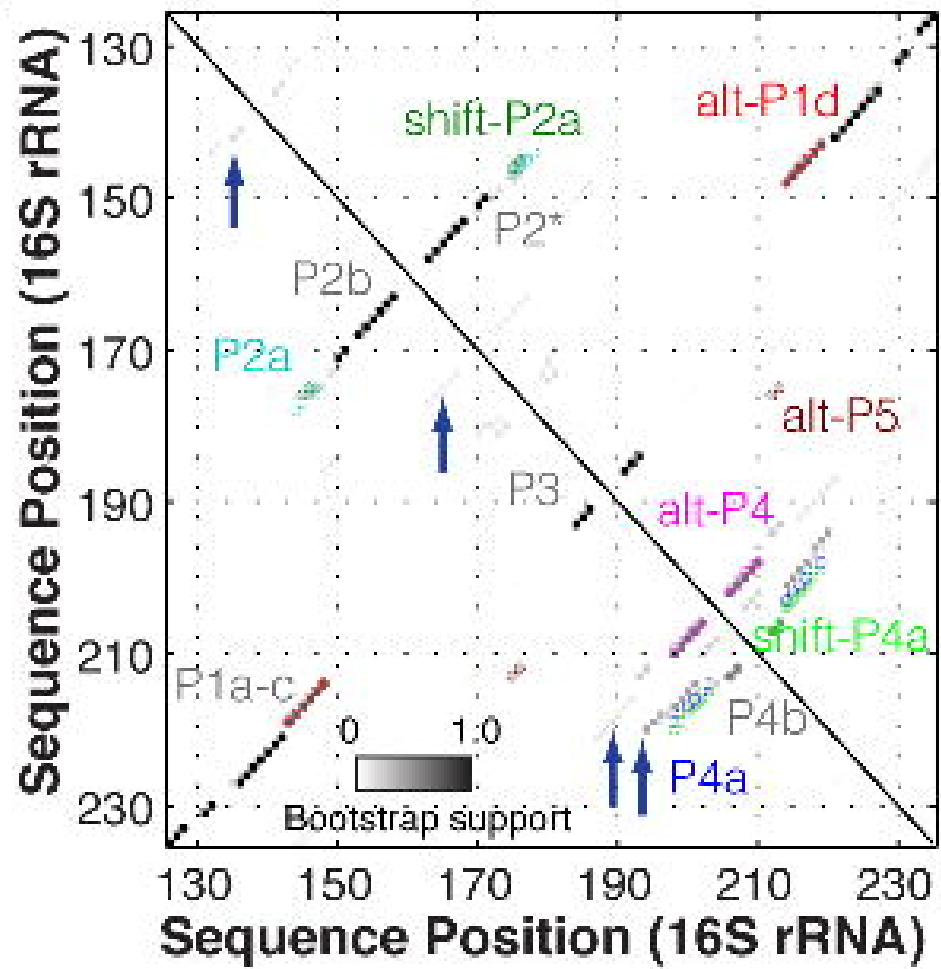

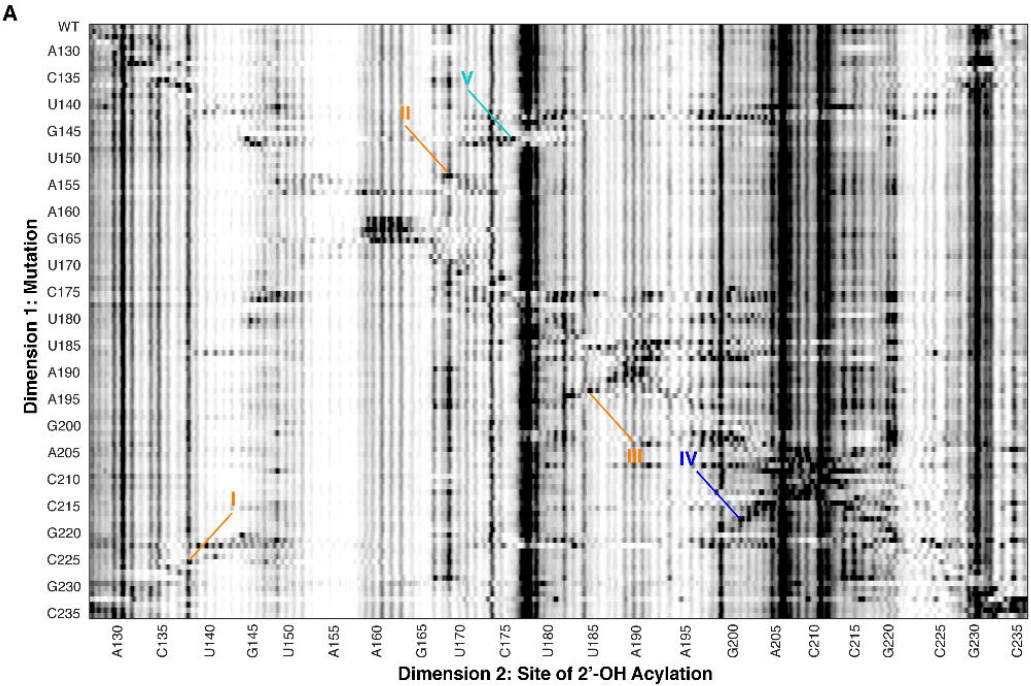
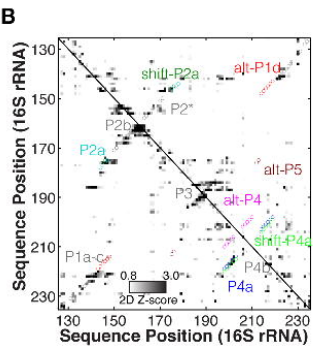
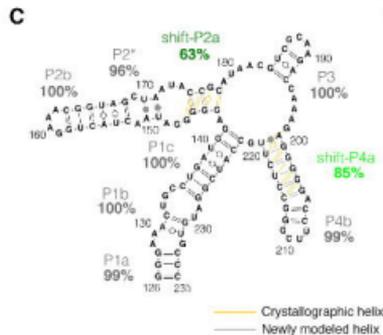
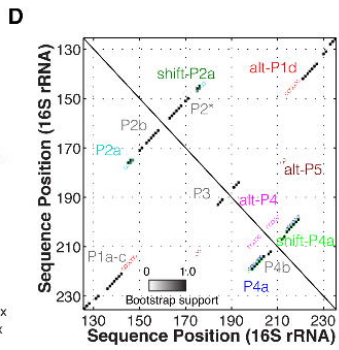

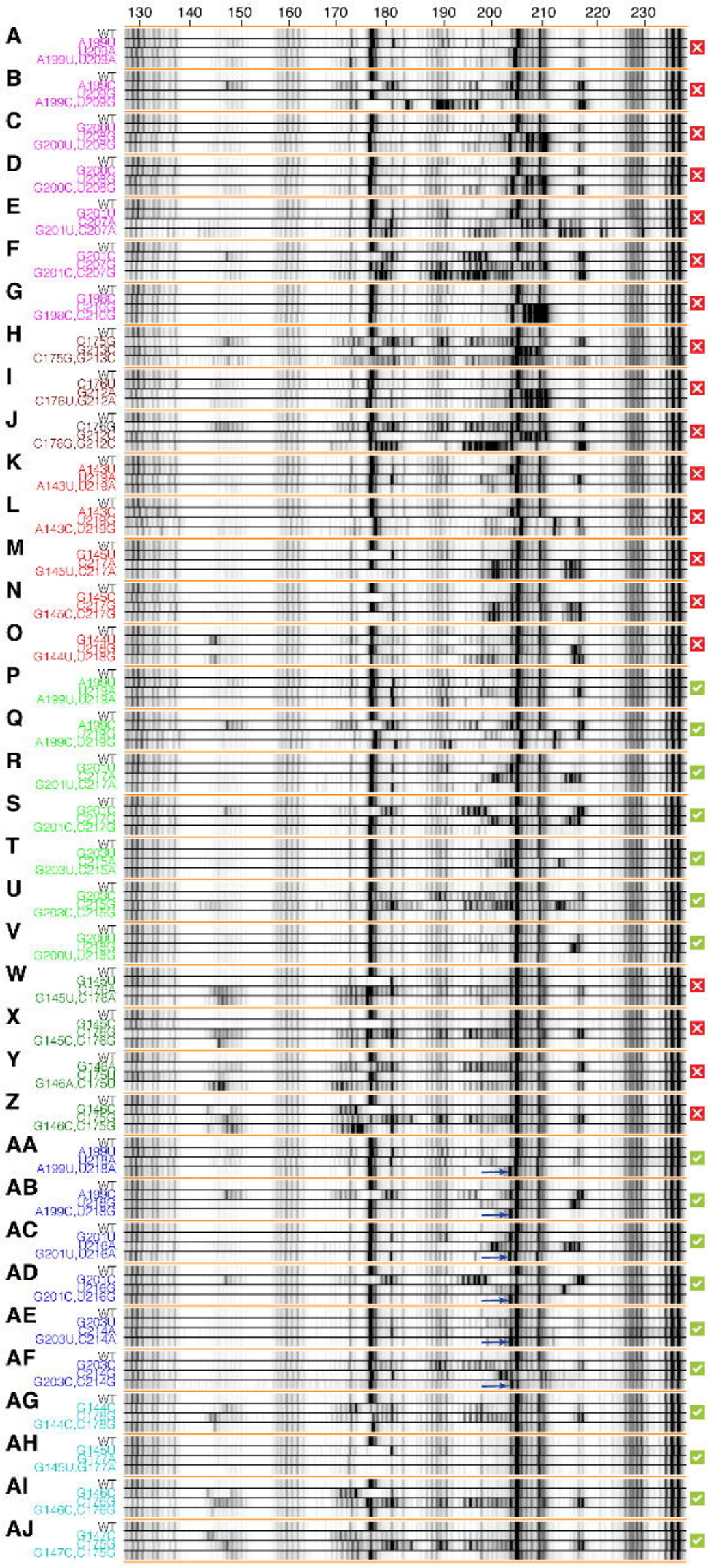

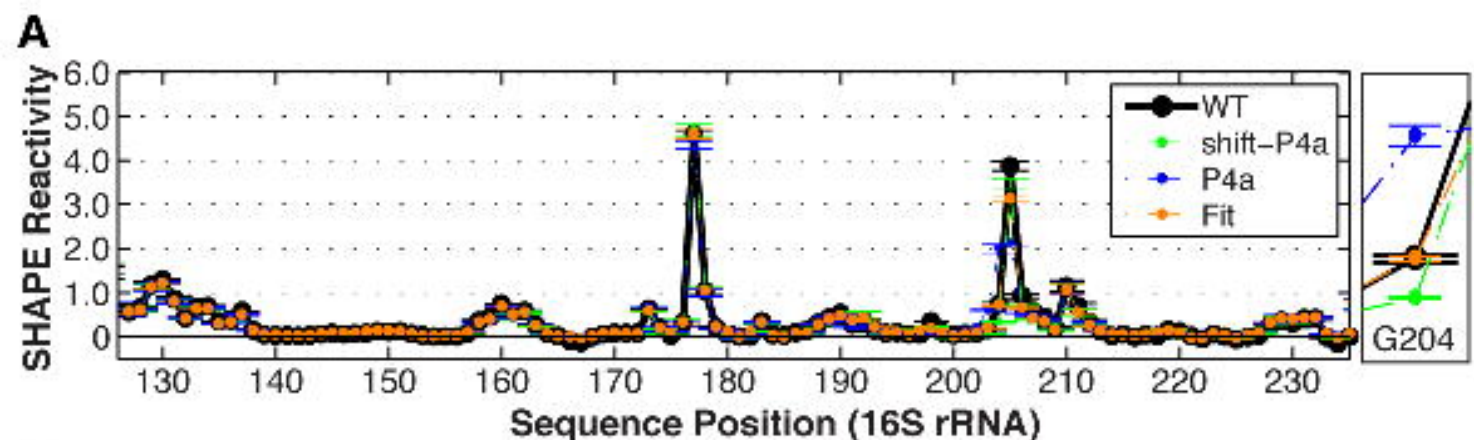